\documentclass[showpacs,draft,preprint,floatfix]{revtex4}
\usepackage{graphicx}
\usepackage{latexsym}
\usepackage{amsmath}
\usepackage{amsfonts}
\usepackage{amssymb}

\begin{document}

\title{On the quantum phase problem}
\author{J.M. Vargas-Mart\'{\i}nez and H. Moya-Cessa}
\affiliation{${}^{1}$ INAOE, Coordinaci\'on de Optica, Apdo.
Postal 51 y 216, 72000 Puebla, Pue., Mexico}
\date{\today}
\begin{abstract}
We present a phase formalism that passes the Barnett-Pegg acid
test, i.e. phase fluctuations for  a number state are the expected
value  $\pi^2/3$ which are the fluctuations for a classical random
phase distribution. The formalism is shown to have consistency
subjected to different approaches.
\end{abstract}
\maketitle

The search for a Hermitean phase operator started with the
beginning of quantum electrodynamics. The problem of phase was
first addressed by Dirac \cite{Dirac} early in the history of
quantum mechanics. However, Dirac's solution was found to suffer
from mathematical difficulties \cite{Louissel}. Since then several
formalisms have been introduced \cite{Susskind,Pegg,Paul} that
however have not been completely satisfactory because the way they
are constructed (see Lynch \cite{Lynch} for a review). One of the
most successful formalisms is that of Pegg and Barnett \cite{Pegg}
that is build in a finite-dimensional Hilbert space and where all
the calculations on physical quantities must be done in such a
space and only after they (the calculations) have been realized
the infinite-dimensional limit is taken. Not surprisingly, the
Pegg-Barnett phase operator passes the Barnett-Pegg "acid test"
\cite{Barnett} that states that the phase fluctuations for a
number state must be equal to $\pi^2/3$, which is the value for a
classical random phase distribution.

Other approaches based on radial integration of quasiprobability
distribution functions have been proposed to describe phase
properties (see for instance \cite{Schleich}). However, (radially)
integrated distributions such as the Wigner function have shown
not to work well because of its negativity \cite{Garraway}.

Other mechanisms to describe phase have been proposed that
directly write the Wigner function not in terms of position and
momentum but on number and phase \cite{Vaccaro}.

Here we would like to put forward a formalism for phase that
passes Barnett-Pegg's acid test.

Classically we may decompose a complex c-number, $A$, in amplitude
and phase by simply writing $A=re^{i\phi}$, with $r=|A|$ and
\begin{equation}
\phi=-i\ln\frac{A}{r}, \label{classical}
\end{equation}
where it is implied that we have chosen the principal branch of
the multi-valued logarithm function.

A Hermitean operator in correspondence to the classical form
(\ref{classical}) was proposed by Arroyo Carrasco and Moya-Cessa
\cite{Arroyo}
\begin{equation}
\hat{\phi}=-\frac{i}{2}\hat{D}(\chi)\left[\ln\left(1+\frac{\hat{a}}{\chi}\right)
-\ln\left(1+\frac{\hat{a}^{\dagger}}{\chi}\right)\right]\hat{D}^{\dagger}(\chi)
, \label{quantum}
\end{equation}
where $\hat{a}$ and $\hat{a}^{\dagger}$ are the annihilation and
creation operator for the harmonic oscillator, respectively,
$\hat{D}(\chi)=e^{\chi(\hat{a}^{\dagger}-\hat{a})}$ is the
displacement operator with $\chi$ a real parameter to ensure
convergence of the series

\begin{equation}
\ln\left(1+\frac{\hat{a}}{\chi}\right) =
\sum_{k=1}^{\infty}\frac{(-1)^{k-1}}{k}\left(\frac{\hat{a}}{\chi}\right)^k.
\end{equation}
The operator (\ref{quantum}) may be found to be Turski's operator
\cite{Turski} by using the unity operator given in terms of
coherent states $ \hat{1}=\frac{1}{\pi} \int
|\alpha\rangle\langle\alpha|d^2\alpha $ and  inserting this
expression into (\ref{quantum}) which yields

\begin{equation}
\hat{\phi}=-\frac{i}{2}\hat{D}(\chi)\left[\ln\left(1+\frac{\hat{a}}{\chi}\right)\frac{1}{\pi}\int
|\alpha\rangle\langle\alpha|d^2\alpha -\frac{1}{\pi}\int
|\alpha\rangle\langle\alpha|d^2\alpha\ln\left(1+\frac{\hat{a}^{\dagger}}{\chi}\right)\right]\hat{D}^{\dagger}(\chi),
\label{turski1}
\end{equation} and that may finally be written as

\begin{equation}
\hat{\phi}=-\frac{i}{2\pi}\int  (\ln\alpha- \ln\alpha^*)
|\alpha\rangle\langle\alpha|d^2\alpha . \label{turski}
\end{equation}

Again, choosing the principal branch in the above equation, we can
rewrite (\ref{turski}) as
\begin{equation}
\hat{\phi}=\frac{1}{\pi}\int  \theta
|\alpha\rangle\langle\alpha|d^2\alpha , \label{theta}
\end{equation}
where $\theta=\arg(\alpha)$. It may be easily shown that the
operator (\ref{theta}) obeys the equation of motion
\begin{equation}
\frac{d\hat{\phi}}{dt}=i\omega[\hat{a}^{\dagger}\hat{a},\hat{\phi}]
\label{evol}
\end{equation}
where $\omega$ is the frequency of the harmonic oscillator. Note
that for a phase operator defined in a finite dimensional Hilbert
space to obey such equation of motion, the harmonic oscillator
Hamiltonian should be defined also in a finite dimensional Hilbert
space \cite{Buzek}.

Dirac's original idea was to define the phase operator as the
argument of the annihilation operator. We can calculate the
average value of the argument of the annihilation operator, given
a wave function $|\psi\rangle$, by using the $Q$-function defined
as

\begin{equation}
Q(\alpha)=\frac{1}{\pi}|\langle\alpha|\psi\rangle|^2,
\end{equation}
by recalling that
\begin{equation}
\langle f(\hat{a})\rangle =\int f(\alpha)Q(\alpha)d^2\alpha
\label{function}
\end{equation}
such that
\begin{equation}
\langle \arg(\hat{a})\rangle =\int \arg(\alpha)Q(\alpha)d^2\alpha,
\label{argument}
\end{equation}
that corresponds to the average value of (\ref{theta}):
\begin{equation}
\langle \psi |\hat{\phi}|\psi\rangle = \frac{1}{\pi}\int \theta
|\langle\alpha|\psi\rangle|^2 d^2\alpha \equiv \langle
\arg(\hat{a})\rangle .
\end{equation}

We can calculate average values of the moments of $\arg(\hat{a})$
via (\ref{function}) as
\begin{equation}
\langle \arg^k(\hat{a})\rangle =\int \theta^k Q(\alpha)d^2\alpha,
\label{argument-k}
\end{equation}
such that we can compute the phase uncertainty, $\Delta \phi =
\langle \arg^2(\hat{a}) \rangle - \langle \arg(\hat{a}) \rangle^2$
for a number state $|n\rangle$, yielding the result
\begin{equation}
\Delta \phi = \frac{\pi^2}{3},
\end{equation}
and where we have used
\begin{equation}
Q(\alpha)= \frac{1}{\pi} |\langle \alpha|n\rangle|^2=
\frac{e^{-|\alpha|^2}}{\pi} \frac{|\alpha|^{2n}}{n!},
\end{equation}
i.e. giving the correct phase uncertainty expected for a state of
undefined phase. We can finally write
\begin{equation}
\arg^k(\hat{a}) \equiv \hat{\phi^k} = \hat{\phi^k} \frac{1}{\pi}
\int |\alpha\rangle\langle\alpha|d^2\alpha \equiv \frac{1}{\pi}
\int \theta^k |\alpha\rangle\langle\alpha|d^2\alpha
\end{equation}
and
\begin{equation}
\hat{e^{i\phi}} = \frac{1}{\pi} \int e^{i\theta}
|\alpha\rangle\langle\alpha|d^2\alpha
\end{equation}
such that
\begin{equation}
 \hat{e^{-i\phi}}
\hat{e^{i\phi}} = \hat{e^{-i\phi}} \frac{1}{\pi} \int e^{i\theta}
|\alpha\rangle\langle\alpha|d^2\alpha=\frac{1}{\pi} \int
e^{-i\theta}e^{i\theta} |\alpha\rangle\langle\alpha|d^2\alpha=
\hat{1}
\end{equation}
i.e. the exponential of phase is a unitary operator within the
formalism.

In conclusion, we have presented a formalism for phase that passes
Barnett-Pegg's acid test giving the correct phase uncertainty for
a number state.

\end{document}